\documentclass[aps,prc,showpacs,showkeys]{revtex4}
\usepackage{graphicx,amsmath,amssymb,bm}

\newcommand{\eq}[1]{Eq.~(\ref{#1})}
\newcommand{\be}{\begin{equation}}
\newcommand{\ee}{\end{equation}}
\newcommand{\bea}{\begin{eqnarray}}
\newcommand{\eea}{\end{eqnarray}}
\newcommand{\bfr}{\bf r}

\def\bfq {{\bf q}}\def\bfb{{\bf b}}
\def\bfr {{\bf r}}\def\bfR {{\bf R}}

\def\bfp{{\bf p}}

\begin{document}

\title{\hskip10cm NT@UW-08-27\\Singular Charge Density at the Center of the Pion?
}
\author{ 
Gerald   A. Miller}

\affiliation{Department of Physics,
University of Washington\\
Seattle, Washington 98195-1560
}

\begin{abstract} We  relate 
the three-dimensional infinite momentum frame spatial charge density
of the pion to its  
electromagnetic form factor $F_\pi(Q^2)$. Diverse treatments of the
measured form factor data including: 
phenomenological fits,  non-relativistic quark models, the application
of perturbative QCD,  QCD sum rules, holographic QCD and the 
Nambu Jona-Lasinio  (NJL) model all
lead to the result that the charge density at the center of the pion has a
logarithmic divergence. Relativistic  constituent 
quark models do not display this  singularity. Future measurements
planned for larger values of $Q^2$  may determine whether or not a
singularity actually occurs. 

\end{abstract}
\maketitle
Understanding the  pion is a necessary step to learning  how QCD
describes the interaction and existence of 
elementary particles.
  As a nearly massless excitation of the QCD vacuum with 
pseudoscalar quantum numbers, the pion  plays a central role in particle and
nuclear physics 
as a harbinger of spontaneous symmetry breaking 
and as   the carrier of the longest range force between nucleons.  

The importance of the pion has been recognized by
a huge level of both experimental and theoretical activity aimed
 at measuring its properties and understanding its structure.
New measurements of the pion electromagnetic form factor, $F_\pi(Q^2)$,
 have been performed \cite{Blok:2008jy,Huber:2008id} and are 
planned \cite{pi12}.  Here we present
 the first 
 phenomenological analysis of existing data to determine the charge density of 
 the pion in a model independent manner.

A proper determination of a 
 charge density requires the measurement of a density operator.
We shall show that measurements of the pion form factor directly involve the 
three-dimensional charge
density of partons, in the infinite momentum frame,  $\hat{\rho}_\infty(x^-,\bfb)$.
In this frame \cite{notation}, the electromagnetic charge density 
$J^0$ becomes $J^+$ and
\bea \hat{\rho}_\infty(x^-,\bfb)=J^+(x^-,\bfb)=\sum_q e_q \overline{q}(x^-,\bfb)\gamma^+q(x^-,\bfb)=\sum_q e_q \sqrt{2} 
q^\dagger_+(x^-,\bfb)q_+(x^-,\bfb),\label{imfop}\eea
%\sqrt{2}\sum_q e_q q_+^\dagger(0)q_+(0) 
%\ee
where %$q(x^\mu)$ are quark field operators, and
 $q_+(x^\mu)=
\gamma^0\gamma^+/\sqrt{2} q(x^\mu)$, the independent part of the
 quark-field operator $q(x^\mu)$.
We set the 
  time variable, $x^+=(t+z)/\sqrt{2},$ to  zero, and do not display
 it in  any function.

We are concerned with the relationship between charge density and 
the  electromagnetic form factor $F_\pi(Q^2)$,  which is determined from the current density via the relation:
\bea F_\pi(Q^2)={\langle {p'}^+,\bfp'|J^+(0)|p^+,\bfp\rangle\over 2p^+},\label{fdef}\eea
where  states are normalized as 
$\langle {p'}^+,\bfp'| {p}^+,\bfp,\rangle
=2p^+(2\pi)^3  \delta({p'}^+-p^+)\delta^{(2)}({\bfp}'-\bfp)$. We take 
 the momentum transfer $q_\alpha=p'_\alpha-p_\alpha$ to be   space-like, with  
$Q^2\equiv -q^2>0,$ and  use  the Drell-Yan (DY) frame with 
$ (q^+=0,Q^2=\bfq^2)$. 
The  matrix element appearing in \eq{fdef} involves the combination
of creation and destruction operators: $b^\dagger b- d^\dagger d$ for each flavor of quark,
so that the valence charge density  is probed. Note also that 
the form factor $F_1$ is independent of
 renormalization scale  because the vector current $\bar{q}\gamma^\mu q$ is conserved \cite{diehl2}. 

The spatial structure of a  hadron can be examined if
one  uses \cite{soper1,mbimpact,diehl2} 
states that are transversely localized.  The state with transverse center of mass
$\bfR$ set to {\bf 0} is formed by taking a  linear superposition of
states of transverse momentum:
\be 
\left|p^+,{\bf R}= {\bf 0},
\lambda\right\rangle
\equiv {\cal N}\int \frac{d^2{\bf p}}{(2\pi)^2} 
\left|p^+,{\bf p}, \lambda \right\rangle,
\label{eq:loc}
\ee
where $\left|p^+,{\bf p}, \lambda \right\rangle$
are light-cone helicity eigenstates
\cite{soper} and
${\cal N}$ is a normalization factor satisfying
$\left|{\cal N}\right|^2\int \frac{d^2{\bf p}_\perp}{(2\pi)^2}=1$.
Wave packet representations can be used to   avoid states 
normalized to $\delta$ functions 
\cite{mb1,diehl},
 but  this  leads to the
same results as using \eq{eq:loc}. Considering that $2p^+p^--\bfp^2=m_\pi^2>0$, one finds that $p^+$ must approach infinity.
This ultra-large value of $p^+$    (infinite momentum frame)  maintains the interpretation
of a pion moving with well-defined longitudinal momentum\cite{mb1}. It is in just such a frame that 
the interpretation of a hadron as a  set of a large number of partons is valid. 
Setting 
the  transverse 
center of momentum of 
a state of  total very large 
momentum $p^+$  to zero as in \eq{eq:loc}, allows
the transverse distance $\bfb$ relative to $\bfR$
to be  defined.

Next  we relate the charge density  
\bea {\rho}_\infty(x^-,\bfb)={ \left\langle p^+,{\bf R}= {\bf 0},
\lambda\right| \hat{\rho}_\infty(x^-,\bfb)
\left|p^+,{\bf R}= {\bf 0},
\lambda\right\rangle
\over 
\left\langle p^+,{\bf R}= {\bf 0},
\lambda|p^+,{\bf R}= {\bf 0},
\lambda\right\rangle},\eea
to $F_\pi(Q^2)$.
In the DY frame no momentum is transferred in the plus-direction, so  that
information regarding the $x^-$ dependence of the distribution is not
accessible. 
Therefore we 
integrate over $x^-$, using the relationship
\be q^\dagger_+(x^-,\bfb)q_+(x^-,\bfb)=
e^{i\widehat{p}^+x^-}e^{-i\widehat{\bfp}\cdot\bfb}q^\dagger_+(0)q_+(0)
e^{i\widehat{\bfp}\cdot\bfb}e^{-i\widehat{p}^+x^-},
\label{trans}\ee to find
\bea
\rho(b)\equiv\int dx^-\rho_\infty(x^-,\bfb)=
\left\langle p^+,{\bf R}= {\bf 0},
\lambda\right| \hat{\rho}_\infty(0,\bfb)
\left|p^+,{\bf R}= {\bf 0},
\lambda\right\rangle/(2p^+). %=\sqrt{2}\sum_qe_q\int dx\;q(x,\bfb), \label{key}
\eea
Furthermore, the use  of Eqs.~(\ref{trans},\ref{eq:loc},\ref{fdef}) leads to  
the simplification of the right-hand-side of the above equation:
\bea
\rho(b)=\int {d^2q\over (2\pi)^2} F_\pi(Q^2=\bfq^2)
e^{-i\bfq\cdot\bfb}, %\qquad{\rm spin \;0}.
\label{rhob0}\eea
where $\rho(b)$ is termed the transverse charge density, giving the charge
density at a transverse position $b$, 
irrespective of the value of the longitudinal position or
momentum.
 This  relation between an integral of
 the three-dimensional infinite momentum frame density and the electromagnetic form factor is
our principal new formula.  Previous results
\cite{soper1,mbimpact,diehl2,Miller:2007uy,Carlson:2007xd} involved the integral
over the longitudinal momentum fraction $x$ of the  impact
parameter parton distribution function (pdf) $q(x,b)$, which  gives the charge density for a quark
at position $b$ for a momentum  fraction (of the plus-component) $x$.
The equality of the respective integrals over $x^-$ or $x$
of the quantities $\rho_\infty(x^-,b)$ and $q(x,b)$ is an example of 
 Parseval's theorem.
The central charge density of the pion is determined by $\rho(b=0)$, because  
the longitudinal dimension  is Lorentz contracted to essentially zero  
in the infinite momentum frame

Recent pion data\cite{Blok:2008jy,Huber:2008id} 
provide an accurate measurement of the 
pion form factor up to a value of $Q^2 $ = 2.45 GeV$^2$. Their
analysis
 includes an assessment of the 
influence of the necessary model dependence caused by extracting the form factor from the
measured cross sections on the experimental error bars.
The existing data for the pion form factor show that it is well represented by the monopole form 
\bea F_\pi(Q^2)=1/(1+R^2Q^2/6),\label{fit}\eea
with $R^2=0.431 \;{\rm fm}^2$. 
A better representation of the data may be a monopole plus dipole \cite{Huber:2008id} 
which involves the square of the term of \eq{fit}, but  any form 
involving
the monopole term leads to a singular central  charge density. This is
because the use 
 \eq{fit} in  \eq{rhob0} gives the result:
\bea \rho(b)=\frac{3 K_0\left(\frac{\sqrt{6} b}{R}\right)}{\pi  R^2},\label{rhobt}\eea
where $K_0$ is modified Bessel function of rank zero. 
For small values of $b$ this function
diverges as $\sim \log(b)$. This divergence is very surprising because the charge density we are considering 
measures a valence quark operator
 between eigenstates of the full Hamiltonian.
The  divergences of  quark distribution
functions that occur at
small values of Bjorken $x$ do not occur here. Any  model, such as
vector meson  dominance or holographic QCD
\cite{Brodsky:2007hb,Kwee:2007dd,Grigoryan:2007wn} that yields
a monopole form factor has a central   density with a
logarithmic divergence..

Intuition regarding a possible singularity in the central
charge density  may be improved by considering other
examples. 
Suppose that  the non-relativistic (NR) limit in which  the quark 
masses are heavy is applicable.
In this case, the pion would be a pure $q\bar{q}$ object and  
the charge density is the Fourier transform of the form factor.
Given the form factor of \eq{fit} the three-dimensional density is uniquely given by
\bea \rho_{\rm NR} (r) =  \frac{3}{2\,\pi \,r\,R^2}\,e^{\frac{{-\sqrt{6}}\,r}{R}}\,\eea
where $r$ is the distance relative to the pion center of mass. If one takes
$r=\sqrt{b^2+z^2}$ as demanded by the rotational invariance of the non-relativistic wave
function, then one finds $\int_{-\infty}^\infty\; dz \rho_{\rm NR}(r)$ is equal to 
$\rho(b) $ of \eq{rhobt}. This is expected because in the NR limit the charge density is 
the same in all frames, including the infinite momentum frame. %is Galilean invariant
%form factor is given by \cite{Brodsky:1989pv}
%\bea F(\bfq^2=Q^2)=\int d^2B dx e^{-i\bfq\cdot\bfB  x}|\psi(B,x)|^2,\eea  
%where $\bfB$ is the transverse distance  between the quark and
%anti-quark, 
%$x$ is the longitudinal momentum fraction carried by the up quark (for
%a $\pi^+$) 
%and the symmetry of the wave function under the interchange 
%$x\leftrightarrow 1-x$ has been used. In the non-relativistic limit
%we make  the replacement
%\bea x\rightarrow {(m+k_3)\over M} , \eea  where $M$ is the pion mass.
%Then $ |\psi(B,x)|^2/M$ is the square of the non-relativistic
%wave-function, 
%$\psi_{NR}$ evaluated at  transverse position $\bfB$ and
%longitudinal position $k_3$, with   canonical position variable $z$. 
%The use of Parceval's theorem gives the non-relativistic form
%factor as
%\bea F(\bfq^2=Q^2)=\int d^2B dz e^{-i\bfq\cdot\bfB m/M}|\psi_{\rm NR}(B,z)|^2.\eea 
%Noting that 
%rotational invariance gives $\psi_{\rm NR}(B,z)=\psi_{\rm
%  NR}(r=\sqrt{B^2+z^2})$ 
%for the lowest energy $l=0$ state allows one to see that 
% the non-relativistic form factor is the three-dimensional 
%Fourier transform of the square of $\psi_{\rm NR}$ or the charge density.
The meaning of the $1/r$ behavior of the density can be understood by considering that
for a $q\bar{q}$ pion, the wave function is the square root of the density so that
 the  short distance wave function $\psi_{\rm NR}\sim
1/\sqrt{r}$. 
Using the Schrodinger equation,
one finds that the potential must contain terms proportional to
$1/r^2$. 
There is no evidence that the  strong interaction potential behaves 
in this manner. Thus  a non-relativistic viewpoint tells us that the
central singularity derived from the form factor falling as $1/Q^2$
requires unsupported assumptions regarding the nature of the
short-distance interactions between quarks. On the other hand, the lowest-energy
solution of the Dirac equation for hydrogenic atoms has a singular
radial  behavior,
$\psi_D(\bfr)\sim1/r^{1-\gamma}e^{-r}$ ($\gamma\equiv \sqrt{1-Z^2\alpha^2}), $
 near the origin at $r=0$. Consider
  $\rho_D(b)\equiv\int_{-\infty}^\infty dz|\psi_D(\bfr)|^2$ % where $r=\sqrt{b^2+z^2}$.
and define  $\eta\equiv 2-2\gamma$, which  ranges between 0 and 2. 
We find for small values of  $b$ (in units of twice the appropriate
Bohr radius) that $\rho_D(b)$ is well behaved for
$0<\eta<1$,  behaves (for all $b$) as $K_0(b) $ for $\eta=1$, and behaves as
$1/b^\eta$ for $1<\eta<2$.  Thus there are physical examples
with a singular central  density.

The  divergence of the central transverse charge density encountered here may be the consequence  of using a simple
parametrization, so we shall consider the predictions of
a variety of different approaches. We begin with
perturbative
QCD (pQCD) 
which provides a prediction \cite{Lepage:1979zb,Efremov:1978rn}
for asymptotically large values of $Q^2$  that
\bea { \lim}_{Q^2\to\infty}F_\pi(Q^2) =16 \pi \alpha_s(Q^2)f_\pi^2/Q^2,\eea  
with the pion decay constant $f_\pi=93 $ MeV, and  in leading order:
\bea  \alpha_s(Q^2)={4\pi\over( 11-{2\over 3}n_f)\ln {Q^2\over
    \Lambda^2}},\label{pqcd}\eea
with $n_f$ the number of quarks of  mass smaller  than $Q$ and $\Lambda$
is a parameter fixed by data. One might think that the $\log Q^2$ term
in the denominator would lead to a non-singular behavior of $\rho(b)$
for small values of $b$. This  is not the case. To see this, consider
the integral:
\bea % {1\over 2\pi}
\int_{Q_0}^{Q_{\rm max}} {dQ\over Q\log Q/\Lambda}J_0(Q b),\eea
for the case $Q_{\rm max}>Q_0>\Lambda>0$. In the limit that $Q_{\rm
  max}$ approaches infinity, this is the contribution of
the integral of \eq{rhob0} arising from values of $Q>Q_0$, assuming that the value of
$Q_0$ is large enough for \eq{pqcd} to be valid. Take $Q_{\rm max}=\epsilon /b$, where $\epsilon$ is a
small positive number such that $J_0(\epsilon)=1$ to any specified
degree of numerical precision. Then
\bea %{1\over 2\pi}
\int_{Q_0}^{\epsilon/b} {dQ\over
    Q\log Q/\Lambda}J_0(Q b)=\log\log({\epsilon\over\Lambda \;b})-\log\log({Q_{0}\over\Lambda}),\;Q_{\rm max}
b\ll1.\label{mid}\eea In  the limit that $b$ approaches zero \eq{mid}
becomes
\bea \lim_{b\to0}%{1\over 2\pi}
\int_{Q_0}^{\infty} {dQ\over
    Q\log Q/\Lambda}J_0(Q b)=\log\log(1/b)+\;\cdots,\label{mid2}\eea
We see that the pQCD form      factor corresponds to a singularity at
short distance. The same feature would arise in any model form factor such as those based on
sum rules {\it e.g} \cite{Radyushkin:1990te}
that  joins smoothly to the pQCD result at very  large values of $Q^2$.

Chiral quark models (see the review   \cite{Broniowski:2008tg})
present other  examples of transverse charge densities that are
singular at the center.
In those models, the pion form factor takes the monopole form
of \eq{fit} so that
 the  central  density diverges as $\log b$ at the origin. Nevertheless  all
physical observables, including $f_\pi$ and structure functions, are
computed to be finite.
We consider two such models. The first is the spectral quark model SQM 
\cite{RuizArriola:2003bs,Broniowski:2003rp}
In this model $F_\pi(Q^2)$  takes the form of \eq{fit} with
$R^2/6=m_\rho^2$. The impact parameter dependent parton  distribution function
is given by  \cite{Broniowski:2003rp}
\bea q(x,b)= { m_\rho^2\over 2\pi(1-x)^2}\left[\frac{-b m_\rho K_1\left(\frac{b m_\rho}{1-x}\right)
   }{(1-x)}+{K_0\left(\frac{b m_\rho}{1-x}\right)}\right]
.\eea
 For small values of $b$ this diverges as $\log b$
for all values of $x$.  Nonetheless, the SQM produces
reasonable structure functions and quark distribution functions
\cite{RuizArriola:2003bs}.

Another example is the NJL model, as regulated  by
two Pauli-Villars subtractions. The form factor is given by  \cite{Broniowski:2008tg}
\bea F_\pi(Q^2)=\int_0^1 dx F_\pi(Q^2,x)\eea
with $
F_\pi(Q^2,x)={-1\over (4\pi)^2}{12 M^2\over f^2}
\ln (M^2 +\Lambda^2+x(1-x)Q^2)_{\rm reg}, $%\eea
where $M$ is the quark mass, $\Lambda$ is a parameter related to
regularization and $f$ is the pion decay constant. The subscript reg
denotes the  regularization procedure \cite{RuizArriola:2002wr}:
$ %\bea
{\cal O}_{\rm reg}(\Lambda^2)={\cal O}(0)-{\cal
  O}(\Lambda^2)+\Lambda^2 {d{\cal O}\over d\Lambda^2}.$ %\eea
The phenomenologically determined  values are $M=280$ MeV, $f$=93.3 MeV, $\Lambda=870 $ MeV.
The impact parameter dependent pdf is the two-dimensional Fourier
transform of  $F_\pi(Q^2,x)$:
\bea
q(x,b)= {-3M^2\over (2\pi)^3 f^2}\int_0^\infty dQ\;QJ_0(Qb)\left[\ln
  \frac{M^2+x(1-x)Q^2}{ M^2+\Lambda^2+x(1-x)Q^2} +\frac{\Lambda^2}{
    M^2+\Lambda^2+x(1-x)Q^2}\right].\eea
This gives a well-behaved expression for $b\to 0$ for all non-zero
values of $x(1-x)$.  Indeed:
\bea q(x,0)= 
 {3M^2\over 2(2\pi)^3 f^2}\left(\Lambda^2+M^2 \log{M^2\over
   M^2+\Lambda^2}\right)\;{1\over x(1-x)}.\eea
Thus a logarithmic divergence appears upon integrating on $x$.

\begin{figure}\label{ffdataandfits}
\unitlength1.cm
\begin{picture}(14,18.2)(-11.5,-.8)
\includegraphics{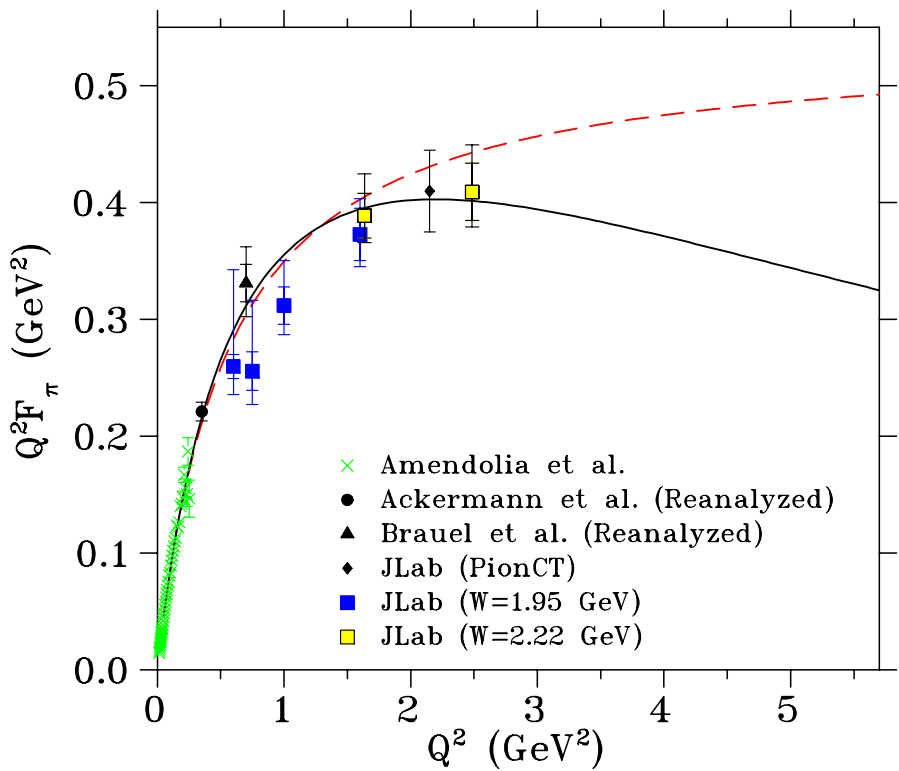}
\end{picture}
\caption{(Color online) $Q^2F_\pi(Q^2)$. Pion form factor data as
  plotted in \cite{Huber:2008id}.  The data labeled Jlab are from
  \cite{Huber:2008id}.
The data  Brauel {\it et al.} \cite{Brauel:1979zk}  and that of
Ackermann {\it et al.} \cite{Ackermann:1977rp} have using the method
  of   \cite{Huber:2008id}. The Amendola data {\it et al.} are from \cite{Amendolia:1986wj}
The data point labeled PionCT is from \cite{PICT}.  The (red) dashed curve uses the monopole fit \eq{fit} and the (black) solid line  the constituent
  quark model of \cite{Hwang:2001hj}.    }% From Huang.nb.}
\end{figure}

Gaussian models with generalized parton distributions $H(x,0,Q^2)$  ($\int dx H(x,0, Q^2)=F(Q^2)$) dominated by behavior
near $x=1$  present a set of examples that also 
 yield a form factor with a $1/Q^2$ asymptotic behavior,
and have a impact parameter distribution that is well behaved at each
value of $x$
 for all $b$. The 
key asymptotic features are captured in the  simple formula 
\cite{Radyushkin:1998rt,Burkardt:2003mb}:
$ %\bea 
H(x,0,Q^2)_{x\to 1}=(1-x)^{n-1}e^{-a(1-x)^n Q^2},\,\quad n>2$
so that %\label{hxq}\\
$q(x,b)_{x\to1}={1\over2\pi a(1-x)}e^{-b^2/(4a(1-x)^n)}$.
This form shows that $q(x,b)$ is well behaved for all values of $b$
and for each value of $x$, but
the integral over $x$ contains a logarithmic divergence. 

\begin{figure}\label{pionrhob}
\unitlength1.cm
\begin{picture}(14,8.2)(1.5,-.8)
\includegraphics{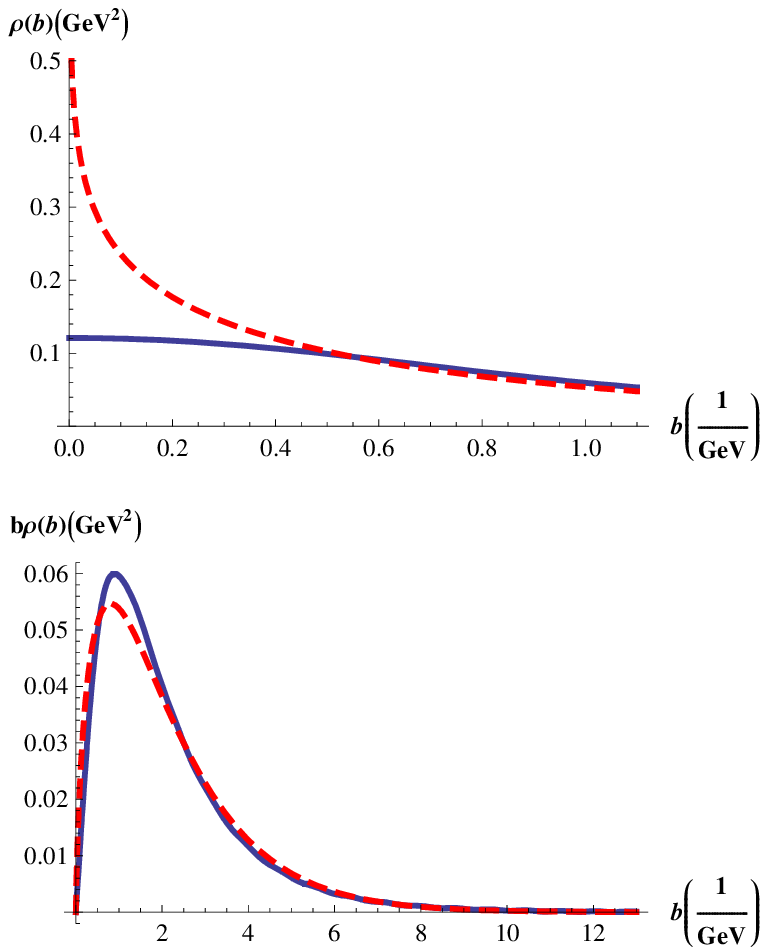}
\end{picture}%\label{rat}
\caption{(Color online) $\rho(b)$ (upper panel) and $b \rho(b)$ (lower panel) corresponding to  the two models
  shown in Fig.~1\ref{ffdataandfits}. The (red) dashed curve uses the monopole fit and the (blue) solid-line the relativistic
constituent
  quark model of\cite{Hwang:2001hj}.  }% From Huang.nb.}
\end{figure}

Not all models that describe the existing form factor data have a
singular central  charge density. Relativistic light-front constituent quark models
\cite{Chung:1988mu,Frederico:1992ye,Hwang:2001hj} are able to describe
the pion phenomenology and the current form factor data. These models produce
a non-singular transverse charge density as we shall illustrate.
These models can be  most simply derived \cite{Frederico:1992ye} by
using the impulse approximation (evaluating  the
triangle diagram).
 One starts by evaluating the integral over the minus component of the  loop momentum $k^\mu$, 
and then cutting off the remaining integral over $x=k^+/p^+,k_\perp$
using a phenomenological wave function that depends on the combination
$(k^2+m^2)/x(1-x)$, with $m$ as the assumed constituent quark mass.  
We illustrate these models by
computing the form factor using the model of \cite{Hwang:2001hj}. The
wave function chosen to be  a power-law form, and the model is able to describe
all of the existing form factor data in both the time-like and space
like regions, $f_\pi$, and the transition form factor $f_{\pi\gamma}$ in
which a virtual photon transforms a real pion into a real photon.
%For space-like momentum transfers the form factor takes the form
The model  form factor of   \cite{Hwang:2001hj} and the monopole fit
of \eq{fit} are shown along with the measured data in
Fig.~1\ref{ffdataandfits}.
Both models provide a good fit to the data, but present very different
predictions for larger values of $Q^2$  where measurements remain to
be done. The corresponding versions of $\rho(b)$ and $b\rho(b)$ are shown in
Fig.~2\ref{pionrhob}.
The singularity contained in \eq{rhobt} appears as a rapidly rising
function as $b$ approaches zero, while the relativistic constituent quark
model provides a $\rho(b)$ that is smooth for small values of $b$.

We summarize.  The high $Q^2$ behavior of the form factor
determines the short distance behavior of the transverse charge density $\rho(b)$. If the form
factor really behaves as the monopole form of \eq{fit}, then
 $\rho(b)$ maintains  a logarithmic
singularity at the origin. A variety of models predict this behavior,
\cite{Brodsky:2007hb,Kwee:2007dd,Grigoryan:2007wn,Broniowski:2008tg,RuizArriola:2003bs,Radyushkin:1998rt,Burkardt:2003mb}
as well as 
 any
non-relativistic constituent quark model that predicts a monopole
behavior of $F_\pi(Q^2)$ and solutions of the Dirac equation.
 
If the form factor falls asymptotically as
perturbative QCD predicts,  the $\rho(b)$ behaves singularly  as
$\ln\ln b$ for small values of $b$.
It seems reasonable that $\rho(b) $, a property of the
valence quark density, should have no singularity. 
The relativistic constituent quark model produces transverse charge
densities that are free of singularities, while providing a 
generally good phenomenology of the pion \cite{Hwang:2001hj}.

It is therefore absolutely and manifestly 
clear that obtaining  data at higher values of
$Q^2$ is essential to providing further understanding. Such 
data could provide support for or rule out either constituent
quark models or current pQCD evaluations of $F_\pi$.  If  an assumption that  the central
density is non-singular is correct,  the form factor will fall as described by 
constituent quark models. On the other hand, 
if asymptotic pQCD
is valid, the central charge density  would be singular-
 a remarkable fact of nature.

%\section*{Acknowledgments}
I thank the USDOE (FG02-97ER41014)
for partial support of this work, and I. Cloet,
 A.~Bernstein, S.~J.~Brodsky, C.~Carlson,  W. Detmold, M. Savage
and L. Wolfenstein for useful discussions. I thank G.~Huber for
conveying the experimental data and the  Physica file  used for  Fig.~1.


\begin{thebibliography}{0}
\bibitem{notation}
Our notation is that $x^\pm\equiv (x^0\pm x^3)/\sqrt{2},p^\pm\equiv (p^0\pm p^3)/\sqrt{2}$, and
$p_\mu x^\mu=p^-x^++p^+x^--\bfp\cdot\bfb$.
The coordinates perpendicular to the 0 and 3 directions are denoted as
$\bfb$ and $\bfp$.
\end{thebibliography}
\end{document}